\documentstyle[12pt,epsfig]{article}

\catcode`@=11
\newcount\@tempcntc
\def\@citex[#1]#2{\if@filesw\immediate\write\@auxout{\string\citation{#2}}\fi
  \@tempcnta\z@\@tempcntb\m@ne\def\@citea{}\@cite{\@for\@citeb:=#2\do
    {\@ifundefined
       {b@\@citeb}{\@citeo\@tempcntb\m@ne\@citea\def\@citea{,}{\bf
?}\@warning
       {Citation `\@citeb' on page \thepage \space undefined}}%
    {\setbox\z@\hbox{\global\@tempcntc0\csname b@\@citeb\endcsname\relax}%
     \ifnum\@tempcntc=\z@ \@citeo\@tempcntb\m@ne
       \@citea\def\@citea{,}\hbox{\csname b@\@citeb\endcsname}%
     \else
      \advance\@tempcntb\@ne
      \ifnum\@tempcntb=\@tempcntc
      \else\advance\@tempcntb\m@ne\@citeo
      \@tempcnta\@tempcntc\@tempcntb\@tempcntc\fi\fi}}\@citeo}{#1}}
\def\@citeo{\ifnum\@tempcnta>\@tempcntb\else\@citea\def\@citea{,}%
  \ifnum\@tempcnta=\@tempcntb\the\@tempcnta\else
   {\advance\@tempcnta\@ne\ifnum\@tempcnta=\@tempcntb \else
\def\@citea{--}\fi
    \advance\@tempcnta\m@ne\the\@tempcnta\@citea\the\@tempcntb}\fi\fi}
\catcode`@=12
\begin{document}
\title{\vskip-3cm{\baselineskip14pt
\centerline{\normalsize DESY 00-107\hfill ISSN 0418-9833}
\centerline{\normalsize hep-ph/0007292\hfill}
\centerline{\normalsize June 2000\hfill}}
\vskip1.5cm
Polarized $J/\psi$ from $\chi_{cJ}$ and $\psi^\prime$ Decays at the Tevatron}
\author{{\sc Bernd A. Kniehl and Jungil Lee}\\
{\normalsize II. Institut f\"ur Theoretische Physik, Universit\"at
Hamburg,}\\
{\normalsize Luruper Chaussee 149, 22761 Hamburg, Germany}}

\date{}

\maketitle

\thispagestyle{empty}

\begin{abstract}
We calculate the partonic cross sections for the hadroproduction of polarized
$\chi_{cJ}$ mesons using the factorization formalism of nonrelativistic QCD.
We also describe how the polarization is passed on to the $J/\psi$ mesons in
the radiative feed-down $\chi_{cJ}\to J/\psi+\gamma$ and in the double-cascade
decays of $\psi^\prime$ mesons via $\chi_{cJ}$ intermediate states.
These represent the missing ingredients needed to predict the polarization of
prompt $J/\psi$ mesons at the Fermilab Tevatron.

\medskip

\end{abstract}

\newpage

\section{Introduction}

Since its discovery in 1974, the $J/\psi$ meson has provided a useful
laboratory for quantitative tests of quantum chromodynamics (QCD) and, in
particular, of the interplay of perturbative and nonperturbative phenomena.
The factorization formalism of nonrelativistic QCD (NRQCD) \cite{bbl} provides
a rigorous theoretical framework for the description of heavy-quarkonium
production and decay.
This formalism implies a separation of short-distance coefficients, which can 
be calculated perturbatively as expansions in the strong-coupling constant
$\alpha_s$, from long-distance matrix elements (ME's), which must be extracted
from experiment.
The relative importance of the latter can be estimated by means of velocity
scaling rules, {\it i.e.} the ME's are predicted to scale with a definite 
power of the heavy-quark ($Q$) velocity $v$ in the limit $v\to 0$.
In this way, the theoretical predictions are organized as double expansions in
$\alpha_s$ and $v$.
A crucial feature of this formalism is that it takes into account the complete
structure of the $Q\overline{Q}$ Fock space, which is spanned by the states
${}^{2S+1}L_J^{(a)}$ with definite spin $S$, orbital angular momentum $L$,
total angular momentum $J$, and colour multiplicity $a=\underline{1}$,
$\underline{8}$.
In particular, this formalism predicts the existence of colour-octet processes
in nature.
This means that $Q\overline{Q}$ pairs are produced at short distances in
colour-octet states and subsequently evolve into physical (colour-singlet)
quarkonia by the nonperturbative emission of soft gluons.
The greatest triumph of this formalism was that it was able to correctly 
describe the cross section of inclusive charmonium hadroproduction measured in
$p\overline{p}$ collisions at the Fermilab Tevatron \cite{abe}, which had
turned out to be more than one order of magnitude in excess of the theoretical
prediction based on the traditional colour-singlet model (CSM) \cite{bai,hum}.

The {\it experimentum crucis} of the NRQCD factorization formalism is the
measurement of the polarization of charmonium with quantum numbers
$J^{PC}=1^{--}$, {\it i.e.} $J/\psi$ and $\psi^\prime$ mesons, in direct
hadroproduction, which is predicted to be fully transverse at sufficiently
large values of the transverse momentum ($p_T$) \cite{cw}.
A convenient measure of the polarization is provided by the variable
\begin{equation}
\alpha=\frac{\sigma_T-2\sigma_L}{\sigma_T+2\sigma_L},
\label{eq:alp}
\end{equation}
where $\sigma_T$ and $\sigma_L$ are the transverse and longitudinal components
of the total cross section $\sigma=\sigma_T+\sigma_L$, respectively.
The unpolarized case is characterized by $\alpha=0$, while $\alpha=1$ and $-1$
correspond to fully transverse and longitudinal polarizations, respectively.
As for the $J/\psi$ and $\psi^\prime$ mesons, $\alpha$ is best measured by
analyzing the angular distributions of their decays to $\mu^+\mu^-$ pairs, 
which are proportional to $1+\alpha\cos^2\theta^*$, where $\theta^*$ is the
angle between the $\mu^+$ direction in the charmonium rest frame and the
charmonium direction in the laboratory frame.
The first such measurement was recently performed for $\psi^\prime$ mesons by
the CDF Collaboration at the Tevatron \cite{aff}.
Unfortunately, the statistics are still too low to allow for a meaningful 
test of the NRQCD prediction \cite{ben,lei}.
In the same experiment, the polarization was also measured for prompt $J/\psi$
mesons, {\it i.e.} those which do not stem from the decays of bottom-flavored
hadrons, with about a hundred times more statistics.
This makes it necessary to include in the theoretical prediction the
production of polarized $J/\psi$ mesons via the decays of the heavier
charmonium states $\chi_{c1}$, $\chi_{c2}$, and $\psi^\prime$, which are
directly produced.
In fact, these feed-down channels make up approximately 15\%, 15\%, and 10\%
of the prompt-$J/\psi$ signal, respectively.
There is an additional, yet smaller contribution to prompt $J/\psi$ production
from the double-cascade decays of directly produced $\psi^\prime$ mesons via
$\chi_{cJ}$ intermediate states.
The purpose of this paper is to provide analytic results for the missing
ingredients which are necessary to make quantitative predictions for all these
contributions.
The resulting prediction for $\alpha$ has already been presented in 
Ref.~\cite{bkl}.
It is consistent with the CDF data at intermediate values of $p_T$, but it
disagrees by about three standard deviations from the data point of largest
$p_T$.

In order to predict $\alpha$ for prompt $J/\psi$ mesons to lowest order (LO)
in the NRQCD factorization formalism, we need the polarized cross sections for
the partonic subprocesses $a+b\to c\overline{c}(n)+d$, where
$a,b,d=g,q,\overline{q}$ with $q=u,d,s$, for $n={}^3S_1^{(1,8)}$,
${}^3P_J^{(1,8)}$, and ${}^1S_0^{(8)}$.
The corresponding results for the unpolarized case are well established 
\cite{bai,hum,cho}.
The polarized cross sections may be found in Refs.~\cite{lei,bkv}, with the
exception of those for the $P$-wave colour-singlet $c\overline{c}$ Fock states
$n={}^3P_1^{(1)}$ and ${}^3P_2^{(1)}$, which are relevant for the $\chi_{c1}$
and $\chi_{c2}$ mesons.
These missing results could have already been obtained in the CSM, before the
advent of the NRQCD factorization formalism.
However, the calculation is somewhat tedious, especially for the spin-two case
of $\chi_{c2}$.
The calculation is further complicated by the fact that $\alpha$ is measured 
in the hadronic center-of-mass (CM) frame, so that one cannot benefit from the
usual simplifications which occur when one chooses the partonic CM frame.
Instead, the cross section does not only depend on the partonic Mandelstam
variables, but also on the boost factor relating the partonic and hadronic
CM frames.
Therefore, the helicity amplitude method, which provides a powerful tool for
the treatment of complicated partonic subprocesses, such as
$g+g\to\chi_{c2}+g$ without polarization, cannot be directly employed.
In Refs.~\cite{ben,lei,bkv}, a Lorentz-covariant formalism for describing the
polarization of massive spin-one particles in the hadronic CM frame was
elaborated.
We adopt this formalism and extend it to the spin-two case.
Furthermore, we need the probabilities for the transitions from the
$c\overline{c}$ Fock states of the $\chi_{c1}$, $\chi_{c2}$, and $\psi^\prime$
mesons to those of the $J/\psi$ meson in the presence of polarization.
Owing to heavy-quark spin symmetry, they are trivial for the single-cascade
decay of the $\psi^\prime$ meson.
The results for the ${}^3P_1^{(1)}$ and ${}^3P_2^{(1)}$ channels may be found
in Ref.~\cite{cwt}, where gluon fragmentation to polarized $\chi_{c1}$ and
$\chi_{c2}$ mesons with subsequent radiative feed-down to polarized $J/\psi$
mesons was studied.
The missing pieces of information are provided here.

This paper is organized as follows.
In Sec.~\ref{sec:vector}, we present a general Lorentz-covariant formalism
which allows us to conveniently describe the partonic cross sections for the
inclusive production of massive spin-one and spin-two bosons, the helicities
of which are fixed in the hadronic CM frame.
In Sec.~\ref{sec:cascade}, we decompose the cross section for the prompt 
production of longitudinal $J/\psi$ mesons into the direct, single-cascade,
and double cascade contributions, and we explain how the various components
factorize into cross sections of partonic subprocesses
$a+b\to c\overline{c}(n)+d$ with definite $c\overline{c}$ helicity, scalar
ME's, and measured branching fractions. 
In Sec.~\ref{sec:alpha}, we present separate theoretical predictions of
$\alpha$ for $J/\psi$ mesons from the single-cascade decays of $\chi_{c1}$,
$\chi_{c2}$, and $\psi^\prime$ mesons, and from the double-cascade decays of
$\psi^\prime$ mesons via $\chi_{cJ}$ intermediate states.
Our conclusions are summarized in Sec.~\ref{sec:discussion}.
In Appendix~\ref{sec:appendix}, we present the ingredients which enter the
analytic expressions for the polarized cross sections of the partonic
subprocesses $a+b\to c\overline{c}(n)+d$, with $n={}^3P_1^{(1)}$ and
${}^3P_2^{(1)}$, which represent the missing links for the theoretical
prediction of $\alpha$ from prompt $J/\psi$ mesons.

\section{Covariant Tensor Decomposition of Polarized Cross Sections}
\label{sec:vector}

The partonic cross sections for the production of massive spin-one and
spin-two bosons, the polarizations of which are to be measured in the hadronic
CM frame, depend not only on the partonic Mandelstam variables, but also on
the boost factor relating the partonic and hadronic CM frames.
A Lorentz-covariant formalism to deal with this problem in the spin-one case
was elaborated in Refs.~\cite{ben,lei,bkv}.
It can be applied to the direct production of polarized $J/\psi$, $\chi_{c1}$,
and $\psi^\prime$ mesons.
In order to also include $\chi_{c2}$ mesons in the polarization analysis, we
need to generalize this formalism to the spin-two case.
Prior to doing that, we recall the spin-one formalism.

In the following, we work in the hadronic CM frame.
Let $\epsilon^\mu(\lambda)$ be the polarization four-vector of a spin-one
boson $H$ with mass $M$, four-momentum $P$, and helicity $\lambda$.
In the case of longitudinal polarization ($\lambda=0$), a Lorentz-covariant
expression reads \cite{ben,bkl,bkv}:
\begin{equation}
\epsilon^\mu(0)=Z^\mu
=\frac{(P\cdot Q/M)P^\mu-MQ^\mu}{\sqrt{(P\cdot Q)^2-M^2s}},
\label{eq:lon}
\end{equation}
where $Q$ is the total four-momentum of the colliding hadrons and $s=Q^2$.
It is convenient to decompose the polarization sum,
\begin{equation}
\rho^{\mu\nu}
=\sum_{\lambda=-1}^1\epsilon^{\mu*}(\lambda)\epsilon^{\nu}(\lambda)
=-g^{\mu\nu}+\frac{P^\mu P^\nu}{M^2},
\end{equation}
into longitudinal and transverse components as
\begin{equation}
\rho^{\mu\nu}=\sum_{|\lambda|=0}^1\rho_{|\lambda|}^{\mu\nu},
\label{eq:dec}
\end{equation}
where
\begin{equation}
\rho_{|\lambda|}^{\mu\nu}
=\sum_{\lambda=\pm|\lambda|}\epsilon^{\mu*}(\lambda)\epsilon^{\nu}(\lambda).
\end{equation}
Specifically, we have
\begin{equation}
\rho_0^{\mu\nu}=Z^\mu Z^\nu,\quad
\rho_1^{\mu\nu}=\rho^{\mu\nu}-\rho_0^{\mu\nu}.
\end{equation}

Let ${\cal M}$ be the transition matrix element of the partonic subprocess
\begin{equation}
a(p_a)+b(p_b)\to H(P,\lambda)+d,
\label{eq:process}
\end{equation}
where the four-momenta and helicities given in parentheses refer to the 
hadronic CM frame.
%, and $p_d=p_a+p_b-P$ due to four-momentum conservation.
If we average the absolute square of ${\cal M}$ over the spin and colour
states of the initial-state partons $a$ and $b$, and sum over 
$\lambda=\pm|\lambda|$ and the spin and colour states of the final-state
parton $d$, we can write the outcome as
\begin{equation}
\overline{|{\cal M}|_{|\lambda|}^2}=S_{\mu\nu}\rho_{|\lambda|}^{\mu\nu},
\label{eq:mat}
\end{equation}
where $S^{\mu\nu}$ is a rank-two Lorentz tensor which depends on $p_a$, $p_b$,
and $P$.
Exploiting the the symmetry and orthogonality properties
$\rho_{|\lambda|}^{\mu\nu}=\rho_{|\lambda|}^{\nu\mu}$ and
$P_\mu\rho_{|\lambda|}^{\mu\nu}=0$, respectively, we can decompose
$S^{\mu\nu}$ as
\begin{equation}
S^{\mu\nu}=F\sum_{i=1}^{4}c_is_i^{\mu\nu},
\label{eq:s1}
\end{equation}
where the common factor $F$ and the coefficients $c_i$ are scalar functions
of the partonic Mandelstam variables
\begin{eqnarray}
\hat s&=&(p_a+p_b)^2,\nonumber\\
\hat t&=&(p_a-P)^2,\nonumber\\
\hat u&=&(p_b-P)^2,
\label{eq:man}
\end{eqnarray}
which satisfy the relation $\hat s+\hat t+\hat u=M^2$, and
\begin{eqnarray}
s_1^{\mu\nu}&=&g^{\mu\nu},\nonumber\\
s_2^{\mu\nu}&=&p_a^\mu p_a^\nu,\nonumber\\
s_3^{\mu\nu}&=&p_b^\mu p_b^\nu,\nonumber\\
s_4^{\mu\nu}&=&p_a^\mu p_b^\nu.
\end{eqnarray}

We now turn to the spin-two case.
Let $\epsilon^{\mu\nu}(\lambda)$ be the polarization tensor of a spin-two
boson $H$ with mass $M$, four-momentum $P$, and helicity $\lambda$.
It can be constructed from the polarization four-vectors of two spin-one
bosons with the same mass and four-momentum with the help of the addition
theorem for two angular momenta as \cite{gub}
\begin{equation}
\epsilon^{\mu\nu}(\lambda)=\sum_{\lambda_1,\lambda_2=-1}^1
\langle1,\lambda_1;1,\lambda_2|2,\lambda\rangle
\epsilon^\mu(\lambda_1)\epsilon^\nu(\lambda_2),
\end{equation}
where $\langle1,\lambda_1;1,\lambda_2|2,\lambda\rangle$ are Clebsch-Gordon
coefficients.
Similarly to Eq.~(\ref{eq:dec}), we decompose the polarization sum,
\begin{equation}
\rho^{\mu\nu\rho\sigma}
=\sum_{\lambda=-2}^2\epsilon^{\mu\nu*}(\lambda)\epsilon^{\rho\sigma}(\lambda)
=\frac{1}{2}\left(\rho^{\mu\rho}\rho^{\nu\sigma}
+\rho^{\mu\sigma}\rho^{\nu\rho}\right)
-\frac{1}{3}\rho^{\mu\nu}\rho^{\rho\sigma},
\end{equation}
as
\begin{equation}
\rho^{\mu\nu\rho\sigma}
=\sum_{|\lambda|=0}^2\rho_{|\lambda|}^{\mu\nu\rho\sigma},
\end{equation}
where
\begin{equation}
\rho_{|\lambda|}^{\mu\nu\rho\sigma}
=\sum_{\lambda=\pm|\lambda|}\epsilon^{\mu\nu*}(\lambda)
\epsilon^{\rho\sigma}(\lambda).
\end{equation}
Specifically, we have \cite{cwt}
\begin{eqnarray}
\rho_0^{\mu\nu\rho\sigma}&=&
\frac{1}{6}\left(2\rho_0^{\mu\nu}-\rho_1^{\mu\nu}\right)
\left(2\rho_0^{\rho\sigma}-\rho_1^{\rho\sigma}\right),\nonumber\\
\rho_1^{\mu\nu\rho\sigma}&=&
\frac{1}{2}\left(\rho_0^{\mu\rho}\rho_1^{\nu\sigma}
+\rho_0^{\mu\sigma}\rho_1^{\nu\rho}+\rho_0^{\nu\rho}\rho_1^{\mu\sigma}
+\rho_0^{\nu\sigma}\rho_1^{\mu\rho}\right),\nonumber\\
\rho_2^{\mu\nu\rho\sigma}&=&
\frac{1}{2}\left(\rho_1^{\mu\rho}\rho_1^{\nu\sigma}
+\rho_1^{\mu\sigma}\rho_1^{\nu\rho}-\rho_1^{\mu\nu}\rho_1^{\rho\sigma}
\right).
\end{eqnarray}

We now consider the case when $H$ in Eq.~(\ref{eq:process}) 
represents a spin-two boson.
Then, Eq.~(\ref{eq:mat}) is replaced by
\begin{equation}
\overline{|{\cal M}|_{|\lambda|}^2}
=S_{\mu\nu\rho\sigma}\rho_{|\lambda|}^{\mu\nu\rho\sigma},
\label{eq:two}
\end{equation}
where $S^{\mu\nu\rho\sigma}$ is a rank-four Lorentz tensor which depends on
$p_a$, $p_b$, and $P$.
Making use of the symmetry, orthogonality, and tracelessness properties
$\rho_{|\lambda|}^{\mu\nu\rho\sigma}=\rho_{|\lambda|}^{\rho\sigma\mu\nu}=
\rho_{|\lambda|}^{\nu\mu\rho\sigma}$,
$P_\mu\rho_{|\lambda|}^{\mu\nu\rho\sigma}=0$, and
$g_{\mu\nu}\rho_{|\lambda|}^{\mu\nu\rho\sigma}=0$, respectively, we can 
constrain the decomposition of $S^{\mu\nu\rho\sigma}$ to be of the form
\begin{equation}
S^{\mu\nu\rho\sigma}=F\sum_{i=1}^{10}c_is_i^{\mu\nu\rho\sigma},
\label{eq:s2}
\end{equation}
where $F$ and $c_i$ are scalar functions of $\hat s$, $\hat t$, and $\hat u$,
defined in Eq.~(\ref{eq:man}), and
\begin{eqnarray}
s_1^{\mu\nu\rho\sigma}&=&g^{\mu\rho}g^{\nu\sigma},\nonumber\\
s_2^{\mu\nu\rho\sigma}&=&g^{\mu\rho}p_a^\nu p_a^\sigma,\nonumber\\
s_3^{\mu\nu\rho\sigma}&=&g^{\mu\rho}p_b^\nu p_b^\sigma,\nonumber\\
s_4^{\mu\nu\rho\sigma}&=&g^{\mu\rho}p_a^\nu p_b^\sigma,\nonumber\\
s_5^{\mu\nu\rho\sigma}&=&p_a^\mu p_a^\nu p_a^\rho p_a^\sigma,\nonumber\\
s_6^{\mu\nu\rho\sigma}&=&p_b^\mu p_b^\nu p_b^\rho p_b^\sigma,\nonumber\\
s_7^{\mu\nu\rho\sigma}&=&p_a^\mu p_a^\nu p_a^\rho p_b^\sigma,\nonumber\\
s_8^{\mu\nu\rho\sigma}&=&p_b^\mu p_b^\nu p_b^\rho p_a^\sigma,\nonumber\\
s_9^{\mu\nu\rho\sigma}&=&p_a^\mu p_a^\nu p_b^\rho p_b^\sigma,\nonumber\\
s_{10}^{\mu\nu\rho\sigma}&=&p_a^\mu p_b^\nu p_a^\rho p_b^\sigma.
\end{eqnarray}

The dependence on the reference frame of the measurement enters through
Eq.~(\ref{eq:lon}) and resides in the factors
$s_{i\mu\nu}\rho_{|\lambda|}^{\mu\nu}$ and
$s_{i\mu\nu\rho\sigma}\rho_{|\lambda|}^{\mu\nu\rho\sigma}$, which are
independent of the partonic subprocess.
These factors explicitly depend on the longitudinal-momentum fractions which
the partons $a$ and $b$ receive from the hadrons from which they spring.
Their calculation is straightforward, and we refrain from listing them here.
The functions $F$ and $c_i$ in Eqs.~(\ref{eq:s1}) and (\ref{eq:s2}) are
independent of the hadron momenta and the reference frame; they only depend on
the partonic subprocess.
The set of these functions which are relevant for the direct hadroproduction of 
polarized ${}^3S_1$ charmonium ($H=J/\psi,\psi^\prime$) in the NRQCD 
factorization formalism may be found in Refs.~\cite{lei,bkv}.
This includes the ${}^3S_1^{(8)}$ channel, which also contributes in the cases
$H=\chi_{c1},\chi_{c2}$.
In Appendix~\ref{sec:appendix}, we provide the missing sets of functions
pertinent to the latter two cases, namely those associated with the
${}^3P_1^{(1)}$ and ${}^3P_2^{(1)}$ channels, respectively.

\section{Polarization in Cascade}
\label{sec:cascade}

We now explain how to calculate the polarization variable $\alpha$, defined in
Eq.~(\ref{eq:alp}), for prompt $J/\psi$ production in the NRQCD factorization 
formalism.
Our approach not only applies to hadroproduction, but also to any other 
production mechanism, such as photoproduction, deep-inelastic scattering,
$e^+e^-$ annihilation, two-photon scattering, {\it etc.}
It is convenient to calculate $\alpha$ from the unpolarized cross section
$\sigma$ and its longitudinal component $\sigma_L$.
Prompt $J/\psi$ mesons originate from direct $J/\psi$ production, from 
single-cascade decays of directly produced $\chi_{cJ}$ and $\psi^\prime$ 
mesons, and from double-cascade decays of directly produced $\psi^\prime$
mesons via $\chi_{cJ}$ intermediate states.
The unpolarized cross section $\sigma^{{\rm prompt}~J/\psi}$ is simply
obtained by adding the unpolarized cross sections of the various
direct-production processes multiplied with the appropriate branching
fractions.

The calculation of the longitudinal cross section
$\sigma_L^{{\rm prompt}~J/\psi}$ is more involved.
In the following, we explain how it can be expressed in terms of the reduced
partonic cross sections $\hat\sigma_{|\lambda|}(n)$ for the production of 
$c\overline{c}$ pairs in Fock states $n$ with definite absolute helicities
$|\lambda|$, scalar ME's, and branching fractions.
We have
\begin{equation}
\sigma_L^{{\rm prompt}~J/\psi}
=\sigma_L^{{\rm direct}~J/\psi}+\sigma_L^{\chi_{cJ}}
+\sigma_L^{\psi^\prime}+\sigma_L^{\psi^\prime\to\chi_{cJ}}.
\label{eq:pro}
\end{equation}
The cross sections for the direct production of polarized $J/\psi$ and 
$\psi^\prime$ mesons is given by
\begin{equation}
\sigma_{L,T}^{{\rm direct}~\psi}=\sum_n\hat\sigma_{0,1}(n)
\left\langle{\cal O}^{\psi}(n)\right\rangle,
\label{eq:dir}
\end{equation}
where the summation is over $n={}^3S_1^{(1)}$, ${}^3S_1^{(8)}$,
${}^1S_0^{(8)}$, and ${}^3P_J^{(8)}$.
The relevant reduced partonic cross sections $\hat\sigma_{0,1}(n)$ may be
extracted from Refs.~\cite{lei,bkv}.
Equation~(\ref{eq:dir}) allows us to obtain the first term in
Eq.~(\ref{eq:pro}).

In the NRQCD factorization formalism, the radiative decay
$\chi_{cJ}\to J/\psi+\gamma$ is approximately treated as an electric dipole
transition.
Due to heavy-quark spin symmetry, the helicity of the $J/\psi$ meson is then
given by the third spin component $m_S$ of the $\chi_{cJ}$ meson.
Thus, angular-momentum conservation constrains the transition probability of a
$\chi_{cJ}$ meson with helicity $m_J$ to a longitudinally polarized $J/\psi$
meson, with $m_S=0$, to be proportional to
$\sum_{m_L=-1}^1|\langle1,m_L;1,0|J,m_J\rangle|^2$.
By applying this kind of analysis to the leading $c\overline{c}$ Fock states
$n={}^3P_J^{(1)}$ and ${}^3S_1^{(8)}$ relevant for the $\chi_{cJ}$ mesons, we
find the second term in Eq.~(\ref{eq:pro}) to be
\begin{eqnarray}
\sigma_L^{\chi_{cJ}}&=&
\left[\frac{1}{3}\hat\sigma_0\left({}^3P_0^{(1)}\right)
\left\langle{\cal O}^{\chi_{c0}}\left({}^3P_0^{(1)}\right)\right\rangle
+\frac{1}{3}\hat\sigma_0\left({}^3S_1^{(8)}\right)
\left\langle{\cal O}^{\chi_{c0}}\left({}^3S_1^{(8)}\right)\right\rangle\right]
B(\chi_{c0}\to J/\psi+\gamma)\nonumber\\
&&{}+\left\{\frac{1}{2}\hat\sigma_1\left({}^3P_1^{(1)}\right)
\left\langle{\cal O}^{\chi_{c0}}\left({}^3P_0^{(1)}\right)\right\rangle
+\left[\frac{1}{2}\hat\sigma_0\left({}^3S_1^{(8)}\right)
+\frac{1}{4}\hat\sigma_1\left({}^3S_1^{(8)}\right)\right]
\left\langle{\cal O}^{\chi_{c0}}\left({}^3S_1^{(8)}\right)\right\rangle
\right\}\nonumber\\
&&{}\times B(\chi_{c1}\to J/\psi+\gamma)
+\left\{\left[\frac{2}{3}\hat\sigma_0\left({}^3P_2^{(1)}\right)
+\frac{1}{2}\hat\sigma_1\left({}^3P_2^{(1)}\right)\right]
\left\langle{\cal O}^{\chi_{c0}}\left({}^3P_0^{(1)}\right)\right\rangle
\right.\nonumber\\
&&{}+\left.
\left[\frac{17}{30}\hat\sigma_0\left({}^3S_1^{(8)}\right)
+\frac{13}{60}\hat\sigma_1\left({}^3S_1^{(8)}\right)\right]
\left\langle{\cal O}^{\chi_{c0}}\left({}^3S_1^{(8)}\right)\right\rangle
\right\}B(\chi_{c2}\to J/\psi+\gamma).
\label{eq:chi}
\end{eqnarray}
Here, we made use of the multiplicity relations
\begin{eqnarray}
\left\langle{\cal O}^{\chi_{cJ}}\left({}^3P_J^{(1)}\right)\right\rangle
&=&(2J+1)
\left\langle{\cal O}^{\chi_{c0}}\left({}^3P_0^{(1)}\right)\right\rangle,
\nonumber\\
\left\langle{\cal O}^{\chi_{cJ}}\left({}^3S_1^{(8)}\right)\right\rangle
&=&(2J+1)
\left\langle{\cal O}^{\chi_{c0}}\left({}^3S_1^{(8)}\right)\right\rangle,
\end{eqnarray}
which follow to leading order in $v^2$ from heavy-quark spin symmetry.
To leading order in $v^2$, the absolute value of the first derivative of the
radial wave function of $P$-wave charmonium at the origin
$\left|R_P^\prime(0)\right|$, which is one of the input parameters of the CSM,
is related to
$\left\langle{\cal O}^{\chi_{c0}}\left({}^3P_0^{(1)}\right)\right\rangle$ by
$\left\langle{\cal O}^{\chi_{c0}}\left({}^3P_0^{(1)}\right)\right\rangle
=3N_c\left|R_P^\prime(0)\right|/(2\pi)$, where $N_c=3$.
The combinatorial factors multiplying the terms in Eq.~(\ref{eq:chi}) which
involve $\hat\sigma_{|\lambda|}\left({}^3P_J^{(1)}\right)$ agree with those 
found in Ref.~\cite{cwt}.

The hadronic decay $\psi^\prime\to J/\psi$ proceeds predominantly through a
double chromoelectric dipole transition.
Since transitions involving a spin flip are suppressed and the recoil momentum 
of the $J/\psi$ meson is negligible, the $J/\psi$ meson has the same helicity
as the $\psi^\prime$ meson.
Thus, the third term in Eq.~(\ref{eq:pro}) may be evaluated as
\begin{equation}
\sigma_L^{\psi^\prime}=\sigma_L^{{\rm direct}~\psi^\prime}
B(\psi^\prime\to J/\psi+X).
\end{equation}

The fourth term in Eq.~(\ref{eq:pro}) may be derived in a similar fashion as
the contribution to $\sigma_L^{\chi_{cJ}}$ from the ${}^3S_1^{(8)}$ channel,
and it reads
\begin{eqnarray}
\sigma_L^{\psi^\prime\to\chi_{cJ}}&=&
\frac{1}{3}\sigma_L^{{\rm direct}~\psi^\prime}
B(\psi^\prime\to\chi_{c0}+\gamma)B(\chi_{c0}\to J/\psi+\gamma)\nonumber\\
&&{}+\left(\frac{1}{2}\sigma_L^{{\rm direct}~\psi^\prime}
+\frac{1}{4}\sigma_T^{{\rm direct}~\psi^\prime}\right)
B(\psi^\prime\to\chi_{c1}+\gamma)B(\chi_{c1}\to J/\psi+\gamma)\nonumber\\
&&{}+\left(\frac{17}{30}\sigma_L^{{\rm direct}~\psi^\prime}
+\frac{13}{60}\sigma_T^{{\rm direct}~\psi^\prime}\right)
B(\psi^\prime\to\chi_{c2}+\gamma)B(\chi_{c2}\to J/\psi+\gamma).
\label{eq:dou}
\end{eqnarray}

\section{Numerical Analysis}
\label{sec:alpha}

We are now in a position to explore the phenomenological implications of the
analytic results presented in Secs.~\ref{sec:vector} and \ref{sec:cascade},
and Appendix~\ref{sec:appendix} for the inclusive hadroproduction of polarized
$J/\psi$ mesons at the Tevatron, with CM energy $\sqrt s=1.8$~TeV.
As in the CDF analysis \cite{aff}, we consider the $p_T$ distribution 
$d\sigma/dp_T$, integrated over the rapidity interval $|y|\le0.6$.
We work at leading order in the NRQCD factorization formalism.
We use the MRST98LO parton density functions (PDF's) of the proton \cite{mar}
and evaluate $\alpha_s(\mu)$ from the one-loop formula with
$\Lambda_{\rm QCD}=174$~MeV \cite{mar}.
We set the renormalization and factorization scales $\mu$ equal to the
transverse mass $m_T=\sqrt{4m_c^2+p_T^2}$, with $m_c=1.5$~GeV.
We adopt the nonperturbative ME's appropriate for our choice of PDF's from 
Table~I of Ref.~\cite{bkl}.
As for the $J/\psi$ and $\psi^\prime$ mesons, fits to the Tevatron data do not
simultaneously constrain
$\left\langle{\cal O}^\psi\left({}^1S_0^{(8)}\right)\right\rangle$ and
$\left\langle{\cal O}^\psi\left({}^3P_0^{(8)}\right)\right\rangle$, but only
the linear combination
$M_r=\left\langle{\cal O}^\psi\left({}^1S_0^{(8)}\right)\right\rangle
+r\left\langle{\cal O}^\psi\left({}^3P_0^{(8)}\right)\right\rangle/m_c^2$,
with some optimal value of $r$.
For simplicity, we assume that
$\left\langle{\cal O}^\psi\left({}^1S_0^{(8)}\right)\right\rangle=
r\left\langle{\cal O}^\psi\left({}^3P_0^{(8)}\right)\right\rangle/m_c^2=
M_r/2$.
We adopt the values of the relevant branching fractions from Ref.~\cite{cas}.
We essentially work within the fusion picture, where the $c\bar c$ bound state
is formed in the primary hard-scattering process.
In the high-$p_T$ regime, only those partonic subprocesses survive where the
$c\bar c$ pair is created from a single gluon which is close to its mass
shell.
These contributions, which contain large logarithms of the form
$\ln(m_T/2m_c)$, are conveniently implemented by using fragmentation functions
(FF's).
These logarithms are then resummed by evolving the FF's from their starting 
scale $\mu=2m_c$ up to the scale $\mu=m_T$.
In the case of the unpolarized cross section $\sigma^{{\rm prompt}~J/\psi}$,
we approximately include this resummation by multiplying the fusion cross
section $\hat\sigma\left({}^3S_1^{(8)}\right)$ with the ratio of the
corresponding fragmentation cross sections with final-state factorization
scales $\mu=m_T$ and $\mu=2m_c$.
In the case of the longitudinal cross section
$\sigma_L^{{\rm prompt}~J/\psi}$, we include the LO fragmentation cross
sections $\hat\sigma_0(n)$ with $n={}^3S_1^{(8)}$, ${}^1S_0^{(8)}$, and
${}^3P_J^{(8)}$ \cite{br}, which are of order $\alpha_s^4$.
These contributions are particularly important because the corresponding 
fusion cross sections vanish to order $\alpha_s^3$.
Further details concerning the implementation of the fragmentation 
contributions may be found in Ref.~\cite{bkl}.

\begin{figure}[ht]
\begin{center}
\epsfig{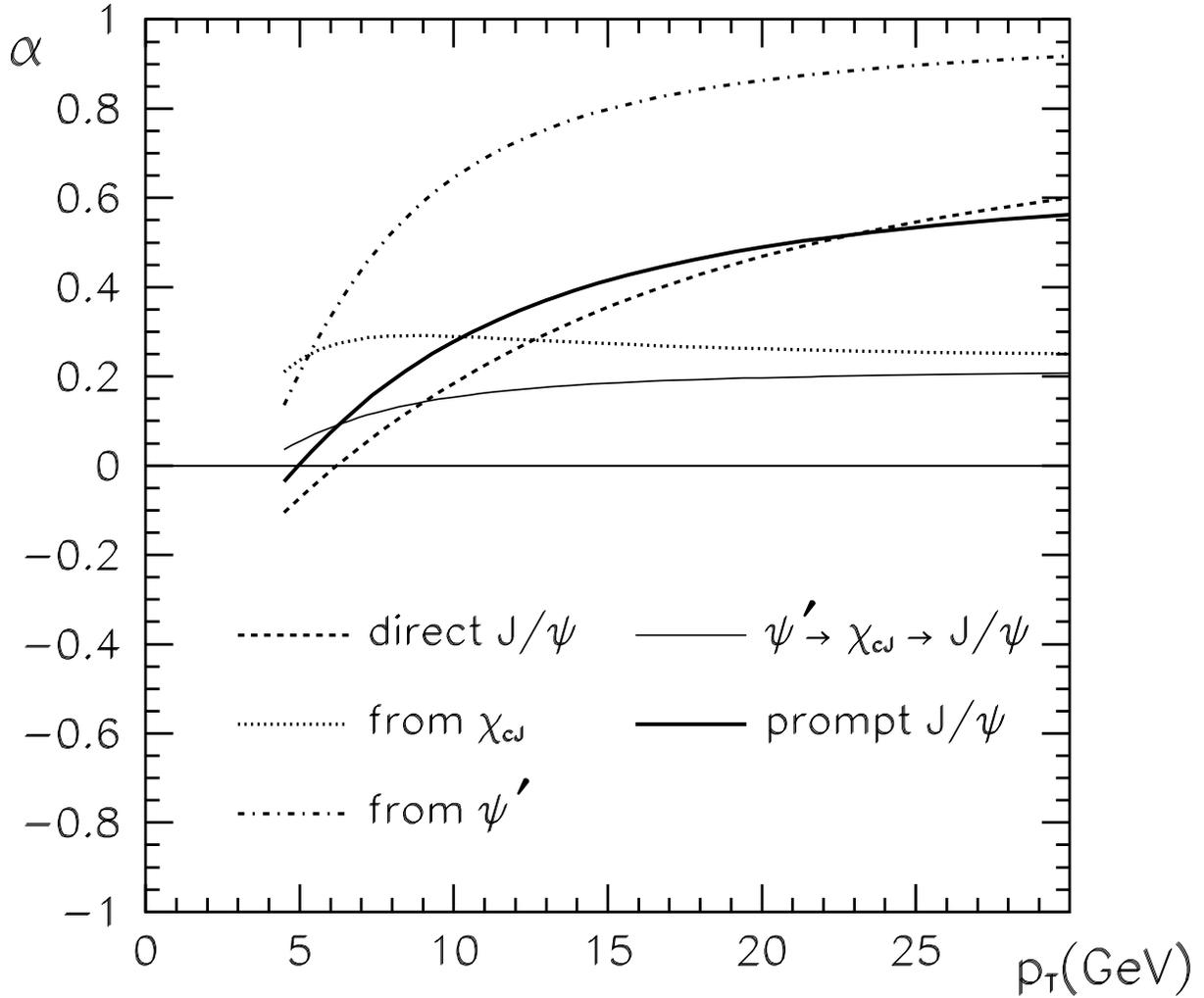}
\end{center}
\caption{Polarization variable $\alpha$ as a function of $p_T$ for direct
$J/\psi$ mesons, for those from single-cascade decays of $\chi_{cJ}$ and
$\psi^\prime$ mesons, for those from double-cascade decays of $\psi^\prime$
mesons via $\chi_{cJ}$ intermediate states, and for prompt ones.}
\label{fig:alp}
\end{figure}

In Fig.~\ref{fig:alp}, we study the $p_T$ dependence of the polarization
variable $\alpha$ for prompt $J/\psi$ mesons.
For comparison, we also show $\alpha$ for the various production mechanisms
which contribute to the prompt-$J/\psi$ signal, namely direct $J/\psi$
production, single-cascade decays of directly produced $\chi_{cJ}$ and
$\psi^\prime$ mesons, and double-cascade decays of directly produced
$\psi^\prime$ mesons via $\chi_{cJ}$ intermediate states.
A quantitative understanding of the asymptotic large-$p_T$ behaviour of
$\alpha$ may be obtained by observing that, in this limit, inclusive 
charmonium production dominantly proceeds via the gluon fragmentation process
$g\to c\overline{c}\left({}^3S_1^{(8)}\right)$ \cite{by}.
Since large-$p_T$ gluons are almost on shell and thus predominantly 
transversely polarized, $\hat\sigma_1\left({}^3S_1^{(8)}\right)$ should 
greatly exceed all other reduced partonic cross sections at large $p_T$ values
\cite{bf}.
It hence follows that, for directly produced $J/\psi$ mesons, $\alpha$
approaches unity in the large-$p_T$ limit \cite{cw}.
By heavy-quark spin symmetry, the same is true for $J/\psi$ mesons from
single-cascade decays of $\psi^\prime$ mesons \cite{bkl}.
The situation is more involved for $J/\psi$ mesons from the radiative
feed-down of $\chi_{cJ}$ mesons and for those from double-cascade decays of
$\psi^\prime$ mesons.
Inspecting Eqs.~(\ref{eq:chi}) and (\ref{eq:dou}) in the large-$p_T$ limit, we
can derive the asymptotic expression
\begin{equation}
\alpha=\frac{60x_0+15x_1+21x_2}{60x_0+75x_1+73x_2},
\end{equation}
where $x_J=(2J+1)B(\chi_{cJ}\to J/\psi+\gamma)$ in the first case and
$x_J=B(\psi^\prime\to\chi_{cJ}+\gamma)B(\chi_{cJ}\to J/\psi+\gamma)$ in the 
second one.
Experimental data suggests that
$B(\chi_{c0}\to J/\psi+\gamma)\ll B(\chi_{c1}\to J/\psi+\gamma)\approx
2B(\chi_{c2}\to J/\psi+\gamma)$ and that $B(\psi^\prime\to\chi_{cJ}+\gamma)$
is approximately independent of $J$ \cite{cas}.
We thus obtain $\alpha^{\chi_{cJ}}\approx39/163\approx0.24$ and
$\alpha^{\psi^\prime\to\chi_{cJ}}\approx51/223\approx0.23$.
(Without these approximations, we find $\alpha^{\chi_{cJ}}\approx0.24$ and
$\alpha^{\psi^\prime\to\chi_{cJ}}\approx0.24$.)
From Fig.~\ref{fig:alp} we see that the corresponding predictions at
$p_T=30$~GeV are already very close to these asymptotic values.
Finally, we observe that the predicted $p_T$ distribution of $\alpha$ from
prompt $J/\psi$ mesons is relatively close to the prediction for direct
$J/\psi$ mesons.
As is evident from Fig.~\ref{fig:alp}, this is a consequence of a strong
cancellation between the $\chi_{cJ}\to J/\psi+\gamma$ and
$\psi^\prime\to J/\psi+X$ channels, which could not have been anticipated
without explicit calculation.

\section{Conclusions}
\label{sec:discussion}

The very recent measurement of the polarization of prompt $J/\psi$ mesons at
the Tevatron \cite{aff} offers a unique opportunity to test the NRQCD
factorization formalism \cite{bbl}, which postulates the existence of 
color-octet processes in nature.
The theoretical prediction of prompt-$J/\psi$ polarization is complicated by
the contributions due to the feed-down from heavier charmonium states, which
come in addition to the direct-$J/\psi$ contribution and make up about 40\% of
the prompt-$J/\psi$ rate.
Specifically, one needs to consider the single-cascade decays of directly
produced $\chi_{c1}$, $\chi_{c2}$, and $\psi^\prime$ mesons, and the
double-cascade decays of directly produced $\psi^\prime$ mesons via
$\chi_{cJ}$ intermediate states.
In this paper, we provided, in analytic form, the missing ingredients which
are necessary to make this prediction.
These include the cross sections of the partonic subprocesses
$a+b\to c\overline{c}(n)+d$ with $n={}^3P_1^{(1)}$ and ${}^3P_2^{(1)}$, and
the probabilities for the transitions from the $c\overline{c}$ Fock states of
the $\chi_{c1}$, $\chi_{c2}$, and $\psi^\prime$ mesons to those of the
$J/\psi$ meson in the presence of polarization.
We presented quantitative predictions, in terms of the polarization variable
$\alpha$, for the various production mechanisms which contribute to the
prompt-$J/\psi$ signal.
In the large-$p_T$ limit, $\alpha$ approaches the asymptotic values
$\alpha^{{\rm direct}~J/\psi}=\alpha^{\psi^\prime}=1$ and
$\alpha^{\chi_{cJ}}\approx\alpha^{\psi^\prime\to\chi_{cJ}}\approx0.24$.

A comprehensive discussion of $\alpha$ for prompt $J/\psi$ mesons together 
with a conservative analysis of the theoretical errors may be found in 
Ref.~\cite{bkl}.
The theoretical prediction \cite{bkl} is consistent with the CDF data 
\cite{aff} at intermediate values of $p_T$, but it exceeds the data point of 
largest $p_T$ by about three standard deviations of the latter.
For the time being, it is premature to conclude that this represents
experimental evidence against the NRQCD factorization formalism.
On the theoretical side, the prediction should be improved by the inclusion of
next-to-leading-order corrections in $\alpha_s$ and $v^2$, which are not yet
available.
This could have an appreciable effect, since
$\alpha_s(2m_c)\approx v^2\approx0.3$ is not actually so small against unity.
On the experimental side, Run~II of the Tevatron is expected to increase the
presently available charmonium data sample by roughly a factor of 20.
This will allow for the polarization to be measured with higher precision and
out to larger values of $p_T$.
Furthermore, it should then be feasible to discriminate between the various
channels which constitute the prompt-$J/\psi$ signal, as is presently done in
the measurement of the unpolarized cross section.
Should the discrepancy persist despite concerted theoretical and experimental
progress, then one possible explanation could be that the NRQCD factorization 
formalism fails for charmonium simply because $m_c$ is too small.
Other effects, such as the one due to nonvanishing intrinsic transverse
momenta of the initial-state gluons \cite{hae}, might be important.
It is, therefore, worthwhile to perform parallel studies for polarized 
bottomonium.

%\bigskip
\newpage
\centerline{\bf ACKNOWLEDGMENTS}
\smallskip

We thank Eric Braaten for his initial contribution, for useful discussions,
and for carefully reading the manuscript.
J.L. thanks Winfried Neun for technical advice concerning programming with
REDUCE.
The work of J.L. was supported by the Alexander von Humboldt Foundation 
through Research Fellowship No.\ IV-KOR/1056268.
This work was supported in part by the Deutsche Forschungsgemeinschaft through
Grant No.\ KN~365/1-1, by the Bundesministerium f\"ur Bildung und Forschung
through Grant No.\ 05~HT9GUA~3, and by the European Commission through the
Research Training Network {\it Quantum Chromodynamics and the Deep Structure
of Elementary Particles} under Contract No.\ ERBFMRX-CT98-0194.

\renewcommand {\theequation}{\Alph{section}.\arabic{equation}}
\begin{appendix}
\setcounter{equation}{0}

\section{\boldmath
Partonic Cross Sections for Color-Singlet Production of Polarized $\chi_{c1}$
and $\chi_{c2}$ Mesons
\unboldmath}
\label{sec:appendix}

In this appendix, we present the ingredients which are necessary to describe 
the direct hadroproduction of polarized $\chi_{c1}$ and $\chi_{c2}$ mesons via
the color-singlet channels $n={}^3P_1^{(1)}$ and ${}^3P_2^{(1)}$,
respectively.
The reduced differential cross sections $d\hat\sigma_{|\lambda|}/d\hat t$ of
the partonic subprocesses $a+b\to c\overline{c}(n)+d$, where the
$c\overline{c}$ pair has absolute helicity $|\lambda|$, are given by
\begin{equation}
\frac{d\hat\sigma_{|\lambda|}}{d\hat t}=\frac{1}{16\pi\hat s^2}
\overline{|{\cal M}|_{|\lambda|}^2}.
\end{equation}
Here, the partons $a$, $b$, and $d$ are treated as massless, while the charm
quark $c$ has mass $m_c$.
In the spin-one (spin-two) case, $\overline{|{\cal M}|_{|\lambda|}^2}$ is 
given by Eq.~(\ref{eq:mat}) (Eq.~(\ref{eq:two})), which contains the
process-dependent tensor $S^{\mu\nu}$ ($S^{\mu\nu\rho\sigma}$).
The functions $F$ and $c_i$ which appear in the Lorentz decomposition of the
latter according to Eq.~(\ref{eq:s1}) (Eq.~(\ref{eq:s2})) are listed below.
As in Eq.~(\ref{eq:chi}), the color-singlet ME
$\left\langle{\cal O}^{\chi_{c0}}\left({}^3P_0^{(1)}\right)\right\rangle$
has been factored out.
Summing over $|\lambda|$, we recover the well-known results for the 
unpolarized case \cite{hum}.
As a further check for our analytical analysis, we also verified the 
differential cross sections of the partonic subprocesses
$a+b\to c\overline{c}\left({}^3P_0^{(1)}\right)+d$, relevant for the 
hadroproduction of $\chi_{c0}$ mesons, given in Ref.~\cite{hum}.

\noindent
$q\overline{q}\to c\overline{c}\left({}^3P_1^{(1)}\right)+g$:
\begin{eqnarray}
F
&=&
-\frac{64(4\pi\alpha_s)^3}{81m_c(\hat{t}+\hat{u})^4},
\nonumber\\
c_1&=&
\hat{t}\hat{u},
\nonumber\\
c_2&=&
2\hat{t},
\nonumber\\
c_3&=&
2\hat{u},
\nonumber\\
c_4&=&
2(\hat{t}+\hat{u}).
\end{eqnarray}

\noindent
$q\overline{q}\to c\overline{c}\left({}^3P_2^{(1)}\right)+g$:
\begin{eqnarray}
F
&=&
\frac{128(4\pi\alpha_s)^3m_c}{81\hat{s}^2(\hat{t}+\hat{u})^4},
\nonumber\\
c_1&=&
\hat{s}(\hat{t}+\hat{u})^2,
\nonumber\\
c_2&=&
4\left[
\hat{s}^2+\hat{t}^2+2\hat{s}(\hat{t}+\hat{u})
\right],
\nonumber\\
c_3&=&
4\left[
\hat{s}^2+\hat{u}^2+2\hat{s}(\hat{t}+\hat{u})
\right],
\nonumber\\
c_4&=&
8\left[
\hat{s}^2-\hat{t}\hat{u}+2\hat{s}(\hat{t}+\hat{u})
\right],
\nonumber\\
c_5&=&c_6=
8\hat{s},
\nonumber\\
c_7&=&c_8=
16\hat{s},
\nonumber\\
c_9&=&
-16(\hat{t}+\hat{u}),
\nonumber\\
c_{10}&=&
16(\hat{s}+\hat{t}+\hat{u}).
\end{eqnarray}

\noindent
$gq\to c\overline{c}\left({}^3P_1^{(1)}\right)+q$:
\begin{eqnarray}
F
&=&
\frac{8(4\pi\alpha_s)^3}{27 m_c(\hat{s}+\hat{u})^4},
\nonumber\\
c_1&=&
\hat{s}\hat{u},
\nonumber\\
c_2&=&
2\hat{s},
\nonumber\\
c_3&=&0,
\nonumber\\
c_4&=&
2(\hat{s}-\hat{u}).
\end{eqnarray}

\noindent
$gq\to c\overline{c}\left({}^3P_2^{(1)}\right)+q$:
\begin{eqnarray}
F
&=&
-\frac{16(4\pi\alpha_s)^3m_c}{27\hat{t}^2(\hat{s}+\hat{u})^4},
\nonumber\\
c_1&=& \hat{t}(\hat{s}+\hat{u})^2,
\nonumber\\
c_2&=& 4\left[\hat{s}^2+\hat{t}^2+2\hat{t}(\hat{s}+\hat{u})\right],
\nonumber\\
c_3&=& 4\hat{s}^2\hat{u}^2,
\nonumber\\
c_4&=& 8 \hat{s}(\hat{s}+\hat{u}),
\nonumber\\
c_5&=&8\hat{t},
\nonumber\\
c_6&=&c_8=0,
\nonumber\\
c_7&=&
16\hat{t},
\nonumber\\
c_9&=&
-16(\hat{s}+\hat{u}),
\nonumber\\
c_{10}&=& 16(\hat{s}+\hat{t}+\hat{u}).
\end{eqnarray}

\noindent
$gg\to c\overline{c}\left({}^3P_1^{(1)}\right)+g$:
\begin{eqnarray}
F
&=&
\frac{(4\pi\alpha_s)^3}
     {6 m_c(\hat{s}+\hat{t})^4(\hat{t}+\hat{u})^4(\hat{u}+\hat{s})^4},
\nonumber\\
c_1&=&
-2\hat{s}^2\hat{t}^2\hat{u}^2(\hat{s}\hat{t}+\hat{t}\hat{u}+\hat{u}\hat{s})
(\hat{s}^2+\hat{t}^2+\hat{u}^2-2\hat{s}\hat{t}-2\hat{t}\hat{u}-2\hat{u}\hat{s})
\nonumber\\
&&{}
- \hat{s}^2\hat{t}^2\hat{u}^2(\hat{s}+\hat{t}+\hat{u})
   \left[\hat{s}\hat{t}(\hat{s}+\hat{t})
              +\hat{t}\hat{u}(\hat{t}+\hat{u})
              +\hat{u}\hat{s}(\hat{u}+\hat{s})\right]
\nonumber\\
&&{}
- (\hat{s}\hat{t}+\hat{t}\hat{u}+\hat{u}\hat{s})
  \left[\hat{s}^3\hat{t}^3(\hat{s}-\hat{t})^2
                      +\hat{t}^3\hat{u}^3(\hat{t}-\hat{u})^2
                      +\hat{u}^3\hat{s}^3(\hat{u}-\hat{s})^2\right],
\nonumber\\
c_2&=&
%-------------------------------------------------------------------
2(\hat{s}+\hat{t})^2
       \left\{
         \hat{s}^2(\hat{t}+\hat{u})
                  \left[
               \hat{s}(\hat{s}-\hat{t})(\hat{s}-\hat{u})(\hat{t}+\hat{u})
                   - \hat{t}\hat{u}(\hat{t}-\hat{u})^2
                  \right]
\right.
\nonumber\\
&&\left.{}
         +(\hat{t}-\hat{u})
          \left[\hat{t}^2+\hat{u}^2-\hat{s}(\hat{t}+\hat{u})\right]
          \left[ \hat{s}^2\hat{t}^2+\hat{t}^2\hat{u}^2+\hat{u}^2\hat{s}^2
           +\hat{s}\hat{t}\hat{u}(\hat{s}+2\hat{t})\right]
       \right\}
%-------------------------------------------------------------------
,
%-------------------------------------------------------------------
\nonumber\\
c_3&=&c_2|_{\hat{t}\leftrightarrow \hat{u}} ,
\nonumber\\
c_4&=&
4(\hat{t}+\hat{u})^2\left\{
      \hat{s}^3(\hat{s}^4+\hat{t}^4+\hat{u}^4)
    - \hat{s}\hat{t}\hat{u}\left[\hat{s}^4
                 +\hat{s}^2(5\hat{t}^2-7\hat{t}\hat{u} +5\hat{u}^2)
                 + \hat{t}\hat{u}(\hat{t}^2-3\hat{t}\hat{u}+\hat{u}^2)\right]
      \right\}
\nonumber\\
&&{}
-4(\hat{t}+\hat{u})
(\hat{t}-\hat{u})^2\left[  \hat{s}^4(\hat{t}+2\hat{u})(2\hat{t}+\hat{u})
                    - \hat{s}^2\hat{t}\hat{u}(\hat{t}^2+\hat{u}^2)
                      + \hat{t}^3\hat{u}^3  \right].
\end{eqnarray}
\noindent
$gg\to c\overline{c}\left({}^3P_2^{(1)}\right)+g$:
\begin{eqnarray}
F&=&
\frac{
8(4\pi\alpha_s)^3m_c
}{ 3(\hat{s}\hat{t}\hat{u})^2(\hat{s}+\hat{t})^4
(\hat{t}+\hat{u})^4(\hat{u}+\hat{s})^4
 },
\nonumber\\
c_1&=&
\hat{s}\hat{t}\hat{u}
(\hat{s}+\hat{t})^2(\hat{t}+\hat{u})^2(\hat{u}+\hat{s})^2
(\hat{s}^2+\hat{t}^2+\hat{u}^2)
(\hat{s}\hat{t}+\hat{t}\hat{u}+\hat{u}\hat{s})^2,
\nonumber\\
c_2&=&2
       (\hat{s}+\hat{t})^2
       \left\{ 
4\hat{u}^8\hat{s}\hat{t}(\hat{s}+\hat{t})^2 
+ \hat{u}^7\hat{s}\hat{t}(\hat{s}+\hat{t})
  (7\hat{s} + 26\hat{s}\hat{t} +7\hat{t}^2)
\right.  \nonumber\\ &&{} \left.  
- \hat{u}^6\left[\hat{s}^6+\hat{t}^6 - 5\hat{s}\hat{t}(\hat{s}^4+\hat{t}^4) 
- 45\hat{s}^2\hat{t}^2(\hat{s}^2+\hat{t}^2) - 87\hat{s}^3\hat{t}^3\right]
\right.  \nonumber\\ &&{} \left.  
+ \hat{u}^5\hat{s}^2\hat{t}^2(\hat{s}+\hat{t})
  (32\hat{s}^2 + 71\hat{s}\hat{t}+32\hat{t}^2) 
+ \hat{u}^4\hat{s}^2\hat{t}^2\left[ 10(\hat{s}^4+\hat{t}^4) 
+ 64\hat{s}\hat{t}(\hat{s}^2+\hat{t}^2) + 105\hat{s}^2\hat{t}^2\right] 
\right.  \nonumber\\ &&{} \left.  
+ \hat{u}^3\hat{s}^3\hat{t}^3
  (\hat{s}+\hat{t})(17\hat{s}^2 + 32\hat{s}\hat{t} + 17\hat{t}^2) 
+ \hat{u}^2\hat{s}^4\hat{t}^4
  (  8\hat{s}^2 + 11\hat{s}\hat{t} +8\hat{t}^2) 
- \hat{u}\hat{s}^5\hat{t}^5(\hat{s} + \hat{t}) 
\right.  \nonumber\\ &&{} \left.  
- \hat{s}^6\hat{t}^6
         \right\},
\nonumber\\
c_3&=&c_2|_{\hat{t}\leftrightarrow \hat{u}},
\nonumber\\
c_4&=&-2
       (\hat{s}+\hat{t})(\hat{s}+\hat{u})
       \left\{
  \hat{s}^6(\hat{t}^2-\hat{u}^2)^2(\hat{t}+2\hat{u})(2\hat{t}+\hat{u})  
- \hat{s}^5\hat{t}^2\hat{u}^2(\hat{t}+\hat{u})
           (3\hat{t}+5\hat{u})(5\hat{t}+3\hat{u})
\right.  \nonumber\\
&&{} \left. 
- \hat{s}^4\hat{t}\hat{u}
 \left[ 3(\hat{t}^6+\hat{u}^6) + 28\hat{t}\hat{u}(\hat{t}^4+\hat{u}^4) 
  + 83\hat{t}^2\hat{u}^2(\hat{t}^2+\hat{u}^2) + 114\hat{t}^3\hat{u}^3 \right] 
\right.  \nonumber\\
&&{} \left.
- \hat{s}^3\hat{t}^2\hat{u}^2
  \left[11(\hat{t}^5+\hat{u}^5) + 53\hat{t}\hat{u}(\hat{t}^3+\hat{u}^3) 
   + 92\hat{t}^2\hat{u}^2(\hat{t}+\hat{u}) \right]             
\right.  \nonumber\\ &&{} \left.  
   - \hat{s}^2\hat{t}^3\hat{u}^3
     (\hat{t}+\hat{u})^2(13\hat{t}^2 + 10\hat{t}\hat{u} + 13\hat{u}^2)
- \hat{s}\hat{t}^4\hat{u}^4
  (\hat{t}+\hat{u})(5\hat{t}^2 + 6\hat{t}\hat{u} + 5\hat{u}^2)
- 2\hat{t}^6\hat{u}^6
        \right\}, 
\nonumber\\
c_5&=&4
       \hat{s}\hat{t}\hat{u}^2(\hat{s}+\hat{t})^2
       (2\hat{s}\hat{t}+\hat{t}\hat{u}+\hat{u}\hat{s})^2
       \left[
 2\hat{u}(\hat{s}+\hat{t}+\hat{u})^2 
 + (\hat{s}+\hat{t})(\hat{s}\hat{t}-\hat{u}^2)
        \right],
\nonumber\\
c_6&=&c_5|_{\hat{t}\leftrightarrow \hat{u}},
\nonumber\\
c_7&=&4
       \hat{s}\hat{t}\hat{u}^2(\hat{s}+\hat{t})
       \left[
   \hat{t}^5(\hat{s} + \hat{u})(5\hat{s}^2 + 12\hat{s}\hat{u} + 5\hat{u}^2)
   + \hat{s}^3\hat{u}^3(3\hat{s}^2 + 4\hat{s}\hat{u} + 3\hat{u}^2)
\right.  \nonumber\\ &&{} \left.  
   + \hat{u}^5\hat{t}(11\hat{s}^2 + 13\hat{s}\hat{t} + 5\hat{t}^2) 
   + \hat{u}^4\hat{t}
    (37\hat{s}^3 + 68\hat{s}^2\hat{t} + 43\hat{s}\hat{t}^2 + 8 \hat{t}^3) 
\right.  \nonumber\\ &&{} \left.  
   + \hat{u}^3\hat{s}\hat{t}
    (43\hat{s}^3 + 132\hat{s}^2\hat{t} + 124\hat{s}\hat{t}^2 + 45\hat{t}^3) 
   + \hat{u}^2\hat{s}^2\hat{t} 
    (19\hat{s}^3 + 110\hat{s}^2\hat{t} + 156\hat{s}\hat{t}^2 + 82\hat{t}^3) 
\right.  \nonumber\\ &&{} \left.  
   + \hat{u}\hat{s}^3\hat{t}^2(31\hat{s}^2 + 83\hat{s}\hat{t} + 61\hat{t}^2) 
   + \hat{s}^4\hat{t}^3(11\hat{s} + 16\hat{t})
        \right],
\nonumber\\
c_8&=&c_7|_{\hat{t}\leftrightarrow \hat{u}},
\nonumber\\
c_9&=&4
       (\hat{s}+\hat{t})(\hat{s}+\hat{u})
       \left\{
 \hat{s}^5 \left[2(\hat{t}^6+\hat{u}^6) 
+ 5\hat{t}\hat{u}(\hat{t}^4+\hat{u}^4) 
+ 4\hat{t}^2\hat{u}^2(\hat{t}^2+ \hat{t}\hat{u} + \hat{u}^2)\right]
\right.  \nonumber\\ &&{} \left.  
+ \hat{s}^4\hat{t}\hat{u}(\hat{t}+\hat{u})\left[3(\hat{t}^4+\hat{u}^4) 
- \hat{t}\hat{u}(\hat{t}+\hat{u})^2 \right] 
- 2\hat{s}^3\hat{t}^2\hat{u}^2\left[ \hat{t}^4+\hat{u}^4 
+ 6\hat{t}\hat{u}(\hat{t}^2+\hat{u}^2) + 9\hat{t}^2\hat{u}^2 \right]
\right.  \nonumber\\ &&{} \left.  
- \hat{s}^2\hat{t}^3\hat{u}^3(\hat{t}+\hat{u}) (\hat{t}^2+\hat{u}^2)
+ 4\hat{s}\hat{t}^4\hat{u}^4(\hat{t}+\hat{u})^2  
+ 2\hat{t}^5\hat{u}^5(\hat{t}+\hat{u})
       \right\}, 
\nonumber\\
c_{10}&=&-4
       \left\{
\hat{s}^7\left[2(\hat{t}^6+\hat{u}^6) 
+ 2\hat{t}\hat{u}(\hat{t}^4+\hat{u}^4) 
- 21\hat{t}^2\hat{u}^2(\hat{t}^2+\hat{u}^2)
- 46\hat{t}^3\hat{u}^3 \right]
\right.  \nonumber\\
&&{} \left.
+\hat{s}^6(\hat{t}+\hat{u})\left[ 2(\hat{t}^6+\hat{u}^6) 
+ 4\hat{t}\hat{u}(\hat{t}^4+\hat{u}^4) 
- 53\hat{t}^2\hat{u}^2(\hat{t}^2+\hat{u}^2) - 134\hat{t}^3\hat{u}^3 \right]
\right.  \nonumber\\
&&{} \left.
+ \hat{s}^5\hat{t}\hat{u}\left[2(\hat{t}^6+\hat{u}^6)
   -43\hat{t}\hat{u}(\hat{t}^4+\hat{u}^4)
   -256\hat{t}^2\hat{u}^2(\hat{t}^2+\hat{u}^2)-428\hat{t}^3\hat{u}^3\right]
\right.  \nonumber\\
&&{} \left.
- \hat{s}^4\hat{t}^2\hat{u}^2(\hat{t}+\hat{u})\left[15(\hat{t}^4+\hat{u}^4) 
+ 136\hat{t}\hat{u}(\hat{t}^2+\hat{u}^2) + 264\hat{t}^2\hat{u}^2 \right]
\right.  \nonumber\\
&&{} \left.
- \hat{s}^3\hat{t}^3\hat{u}^3(\hat{t}+\hat{u})^2
 (32\hat{t}^2 + 93\hat{t}\hat{u} + 32\hat{u}^2) 
- \hat{s}^2\hat{t}^4\hat{u}^4(\hat{t} + \hat{u})
 (19\hat{t}^2 + 34\hat{t}\hat{u} + 19\hat{u}^2)
\right.  \nonumber\\
&&{} \left.
+ 2\hat{t}^6\hat{u}^6(\hat{s}+\hat{t} + \hat{u}) 
       \right\}. 
\end{eqnarray}

\end{appendix}

\end{document}